\documentclass[a4paper,fleqn,usenatbib,useAMS]{mnras}

\usepackage{color}
\usepackage{graphicx}
\usepackage{amsmath}	
\usepackage{amssymb}	
\usepackage{multicol}        
\usepackage{bm}		
\usepackage{pdflscape}	
\usepackage[caption=false]{subfig}

\usepackage[T1]{fontenc}
\usepackage{ae,aecompl}
\usepackage{enumitem}
\usepackage{colortbl}

\usepackage{newtxtext,newtxmath}

\usepackage{etoolbox}
\makeatletter
\patchcmd\@combinedblfloats{\box\@outputbox}{\unvbox\@outputbox}{}{\errmessage{\noexpand patch failed}}
\makeatother

\title[Cusp-to-core transition in low-mass dwarf galaxies]{Cusp-to-core transition in low-mass dwarf galaxies induced by dynamical heating of cold dark matter by primordial black holes}

\author[P. Boldrini et al.]
{Pierre Boldrini$^{1}$\thanks{Contact e-mail:\href{mailto:boldrini@iap.fr}{boldrini@iap.fr}}, 
Yohei Miki$^{2}$, 
Alexander Y. Wagner$^{3}$, 
Roya Mohayaee$^{1}$, 
Joseph Silk,$^{1,4,5}$
\newauthor
Alexandre Arbey$^{6}$  
\\
$^{1}$Sorbonne Universit\'e, CNRS, UMR 7095, Institut d'Astrophysique de Paris, 98 bis bd Arago, 75014 Paris, France\\
$^{2}$Information Technology Center, University of Tokyo, 5-1-5 Kashiwanoha, Chiba 277-8589, Japan\\
$^{3}$University of Tsukuba, Center for Computional Sciences, Tennodai 1-1-1, Tsukuba, Ibaraki, Japan\\
$^{4}$Department of Physics and Astronomy, The Johns Hopkins University, Baltimore MD 21218, USA\\
$^{5}$Beecroft Institute for Particle Astrophysics and Cosmology, Department of Physics, University of Oxford, Oxford OX1 3RH, UK\\
$^{6}$Univ Lyon, Univ Lyon 1, CNRS/IN2P3, Institut de Physique Nucl\'eaire de
Lyon, UMR5822, F-69622 Villeurbanne, France
}
\date{Last updated 2019 Mai 31; in original form xxxxxxx}

\pubyear{2019}

\DeclareMathOperator{\erf}{erf}
\begin{document}
\label{firstpage}
\pagerange{\pageref{firstpage}--\pageref{lastpage}}
\maketitle

\begin{abstract}
We performed a series of high-resolution $N$-body simulations to examine whether dark matter candidates in the form of primordial black holes (PBHs) can solve the cusp-core problem in low-mass dwarf galaxies. If some fraction of the dark matter in low-mass dwarf galaxies consists of PBHs and the rest is cold dark matter, dynamical heating of the cold dark matter by the PBHs induces a cusp-to-core transition in the total dark matter profile. The mechanism works for PBHs in the 25-100 M$_{\sun}$ mass window, consistent with the LIGO detections, but requires a lower limit on the PBH mass fraction of 1$\%$ of the total dwarf galaxy dark matter content. The cusp-to-core transition time-scale is between 1 and 8 Gyr. This time-scale is also a constant multiple of the relaxation time between cold dark matter particles and PBHs, which depends on the mass, the mass fraction and the scale radius of the initial density profile of PBHs. We conclude that dark matter cores occur naturally in halos comprised of cold dark matter and PBHs, without the need to invoke baryonic processes.
\end{abstract}





\section{Introduction}

The nature of dark matter (DM) is one of the major unsolved problems in astrophysics. The most popular dark matter candidates include  weakly interacting massive elementary particles (WIMPs), such as super-symmetric neutralinos or axions \citep{1996PhR...267..195J,2000RPPh...63..793B,2005PhR...405..279B}. An alternative proposal to explain the nature of dark matter is that DM could be made of macroscopic compact halo objects (MACHOs) such as primordial black holes (PBHs) (\cite{1967SvA....10..602Z,1971MNRAS.152...75H} and more recently  \cite{2010RAA....10..495K,2018PDU....22..137C}). These PBHs could naturally be produced in the early Universe via cosmic inflation, without the need to appeal to new physics beyond the standard model \citep{2017PhRvD..96d3504I,2015PhRvD..92b3524C}. There are currently three allowed mass windows around $4\times10^{-17}$, $2\times10^{-14}$ and 25 - 100 M$_{\sun}$ \citep{2017PhRvD..96b3514C}. PBHs can constitute much or even all of the  dark matter in these mass windows by considering only the most well-established bounds and neglecting those that depend on additional astrophysical assumptions. However, taking into account all of the astrophysical constraints means that the PBH+CDM fraction can still be as much as $\sim 0.1$ \citep{2017PhRvD..96b3514C}, and even larger in the lowest mass windows. Despite  the fact that there is still no direct evidence for PBHs, the 25 - 100 M$_{\sun}$ mass window is of special interest in view of the recent detection of black-hole mergers by LIGO \citep{2016PhRvL.116f1102A}. Moreover, the observed LIGO detection rates can be explained for a PBH mass fraction of order 0.001 to 0.01\citep{2016PhRvL.117f1101S,2017PhRvL.119m1301K}. In the future, the Laser Interferometer Space Antenna (LISA) could potentially also detect PBHs in this mass window \citep{2017arXiv170200786A}. 

Although the cold dark matter paradigm can successfully explain various observations at different scales, there are unresolved problems, most notably  the cusp-core problem  but  also tensions  with overpredictions of dwarf galaxy numbers and the too-big-to-fail and diversity issues (see e.g. \cite{2017ARA&A..55..343B} for a recent review). Here we focus on the core/cusp problem. Measurements of galaxy rotation curves and dynamical models of dwarf spheroidal galaxies (dSphs) have revealed that the density profile of the DM halos is approximately constant at the centres of most dwarf galaxies and  corresponds to a cored profile \citep{1994Natur.370..629M,1995ApJ...447L..25B,2001ApJ...552L..23D,2003ApJ...583..732S,2005AJ....129.2119S,2011ApJ...742...20W}. In contrast, cosmological simulations have generally predicted a steep power-law mass-density distribution at the centres of CDM halos, specifically a cuspy profile \citep{1997ApJ...490..493N,1997ApJ...477L...9F,1998ApJ...499L...5M,2010MNRAS.402...21N}. One explanation for the cusp-core problem within the paradigm of cold dark matter relies on the effect of baryonic physics in converting the cusp into a core via changes in the gravitational potential caused by stellar feedback redistributing gas clouds, generating bulk motions and galactic winds along with heating by dynamical friction of massive clumps \citep[e.g.][]{1996ApJ...462..563N,2011ApJ...736L...2O,2013MNRAS.429.3068T,2012MNRAS.421.3464P,2001ApJ...560..636E,2010ApJ...725.1707G,2011MNRAS.418.2527I}.

In collisionless systems, stellar particle encounters lead to the relaxation of particles with similar kinetic energy and drive the system to energy equipartition. Systems comprised of particles of differing masses will also drive mass segregation processes. As an example, massive stars or MACHOs fall towards the center of the potential well and their energy is transferred to the lighter stars, which move away from the center \citep{1969ApJ...158L.139S,1943RvMP...15....1C}. Consequently, the system can expand and the density profile of the system can change due to this diffusion process \citep{2016ApJ...824L..31B,2017PhRvL.119d1102K,2018MNRAS.476....2Z}. 

Here, in the numerical experiments, we explore the consequences of assuming the dark matter in galaxies consists of both cold dark matter (CDM) and PBHs. We propose that PBHs, as DM candidates, can induce a cusp-to-core transition in PBH+CDM halos through gravitational heating from two principal mechanisms, dynamical friction by CDM particles on PBHs and two-body relaxation between PBH and CDM. We explore this transition using high performance $N$-body simulations on GPU to probe the PBH+CDM mass fraction $f_\mathrm{m}$ in $10^7$ M$_{\sun}$ dwarf galaxies. Our simulations allow a mass resolution of 1 M$_{\sun}$ for DM particles. We work with PBHs in the 25-100 M$_{\sun}$ mass window, which is consistent with the LIGO detections. The paper is organized as follows. Section 2 provides a description of $N$-body modelling and our numerical simulations. In Section 3, we show our simulation results and discuss the implications of PBH as a dark matter candidate. Section 4 presents our conclusions. 

\section{$N$-body modelling}

In this work, we consider that a fraction of the DM consists of 25-100 M$_{\sun}$ PBHs. Then, a DM halo is composed of CDM particles and PBHs (DM = PBH + CDM). We define the PBH+CDM mass fraction as 
\begin{equation}
f_\mathrm{m} = \frac{M_{\mathrm{PBH}}}{M_{\mathrm{CDM}}},
\label{hop}
\end{equation}
where $M_{\mathrm{PBH}}$ and $M_{\mathrm{CDM}}$ are the total masses of PBHs and CDM particles. For our halo, we assume the NFW form \citep{1996ApJ...462..563N}: 
\begin{equation}
\rho(r) = \rho_{0}\left(\frac{r}{r_{\mathrm{s}}}\right)^{-1}\left(1+\frac{r}{r_{\mathrm{s}}}\right)^{-2},
\end{equation}
with  scale density $\rho_{0}$ and scale length $r_{\mathrm{s}}$. The relaxation time is proportional to $N/\ln(N)$, where $N$ is the number of particles. Assuming 100 M$_{\sun}$ PBHs and $10^7$ M$_{\sun}$ halo, the particle relaxation will take between 1 and 12 Gyr depending on $f_\mathrm{m}$ considered between 1 and 0.01. In comparison, the relaxation time for $10^8$ ($10^9$ M$_{\sun}$) halos are 10 (100) times longer than for $10^7$ M$_{\sun}$ and longer than the age of the Universe, which is why we focus here on a $10^7$ M$_{\sun}$ dwarf galaxy, starting at redshift $z=2$ in the simulations. Given the halo mass and redshift, the halo concentration $c_{200}$ can be estimated from cosmological $N$-body simulations \citep{2012MNRAS.423.3018P}. Our halo is composed of DM and PBH particles with a total mass of $10^{7}$ M$_{\sun}$.

To generate our NFW halos, we use the initial-condition generator, \textsc{magi}. Adoption of a distribution-function-based method ensures that the final realization of the halo is in dynamical equilibrium \citep{2018MNRAS.475.2269M}. We perform our simulations with the high performance collisionless N-body code, \textsc{gothic} \citep{2017NewA...52...65M, Miki2019}. This gravitational octree code runs entirely on GPU and is accelerated by the use of hierarchical time steps in which a group of particles has the same time step. 
We have performed the simulations using NVIDIA GeForce GTX 1080 Ti (CUDA 9.0) on Intel Xeon Silver 4114 (GCC 4.8). We evolve different PBH+CDM halos composed of 25, 50, 75 and 100 M$_{\sun}$ PBHs over 11 Gyr by adopting the softening length of $\epsilon_0=$1.331 pc and the accuracy control parameter of $2^{-7}$. All runs were made with CDM particles of 1 M$_{\sun}$. We explored also PBH+CDM halos with different mass fractions $f_\mathrm{m}$ = $[0.01,0.1,0.5]$. In our simulations, we assume that the CDM and PBH components of PBH+CDM halos initially follow NFW profiles with the same concentration. We tested two different scale lengths for the profile of the PBH component: $r_{\mathrm{s}}^{\mathrm{PBH}} = r_{\mathrm{s}}^{\mathrm{CDM}}$ and $r_{\mathrm{s}}^{\mathrm{PBH}} = r_{\mathrm{s}}^{\mathrm{CDM}}/2$. The second corresponds to a scenario where the density of PBHs is enhanced in the central region. Indeed, we suppose that mass segregation of PBHs would occur, and  this increases the density of PBHs at the centre. This scenario should also enhance the formation of cores due to dynamical heating of the CDM by PBHs. It will also accelerate the two-body relaxation. 

\section{Results}

\subsection{Evidences for core formation from gravitational heating by PBHs}

\begin{figure}
\centering
\includegraphics[width=0.47\textwidth]{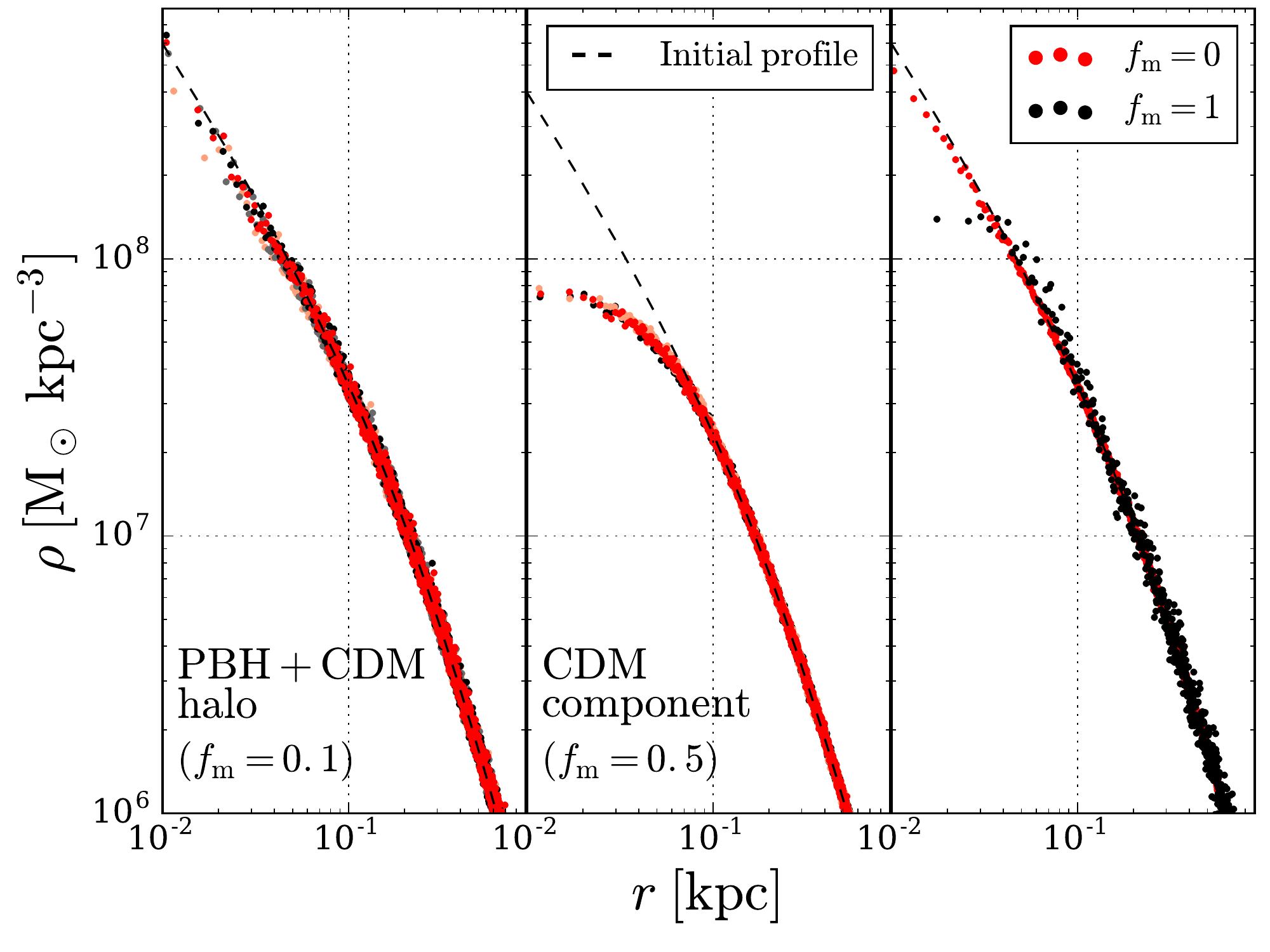}
\caption{{\it Tests of numerical accuracy:} Density profiles of a halo composed of 100 M$_{\sun}$ PBHs to test the impact of the accuracy control parameter A ({\it left panel}) and softening length $\epsilon$ ({\it middle panel}) in simulations. We tested the impact of the accuracy control parameters of $2^{-6}$ (orange points), $2^{-7}$ (black points), $2^{-8}$ (grey points) and $2^{-9}$ (red points) for a halo with $f_\mathrm{m}=0.1$ and $r_{\mathrm{s}}^{\mathrm{PBH}} = r_{\mathrm{s}}^{\mathrm{CDM}}$. Density profiles of PBH+CDM halo ({\it left panel}) hold the initial distribution for all the different parameter values. The middle panel shows our softening convergence test for the density profiles of the CDM component of $f_\mathrm{m}$=0.5 halo with a softening length $\epsilon=\epsilon_0/2$, $\epsilon_0$ and $2\epsilon_0$, where $\epsilon_0$=1.331 pc. Indeed, the core size of the CDM component is independent of the softening length. The right panel describes density profiles of a halo composed only with CDM particles ($f_\mathrm{m}$=0) and a halo composed only with PBH particles ($f_\mathrm{m}$=1). The CDM profiles at the beginning $T=0$ Gyr (dashed line) and at the end of the simulation $T=11$ Gyr (red points), which are nearly identical, show the stability of our halo. However, there is core formation for $f_\mathrm{m}$=1 due only to two-body relaxation between PBHs.}
\label{fga1}
\end{figure}

\begin{figure*}
\centering
\begin{minipage}[t]{8.4cm}
\centering 
\includegraphics[width=8cm]{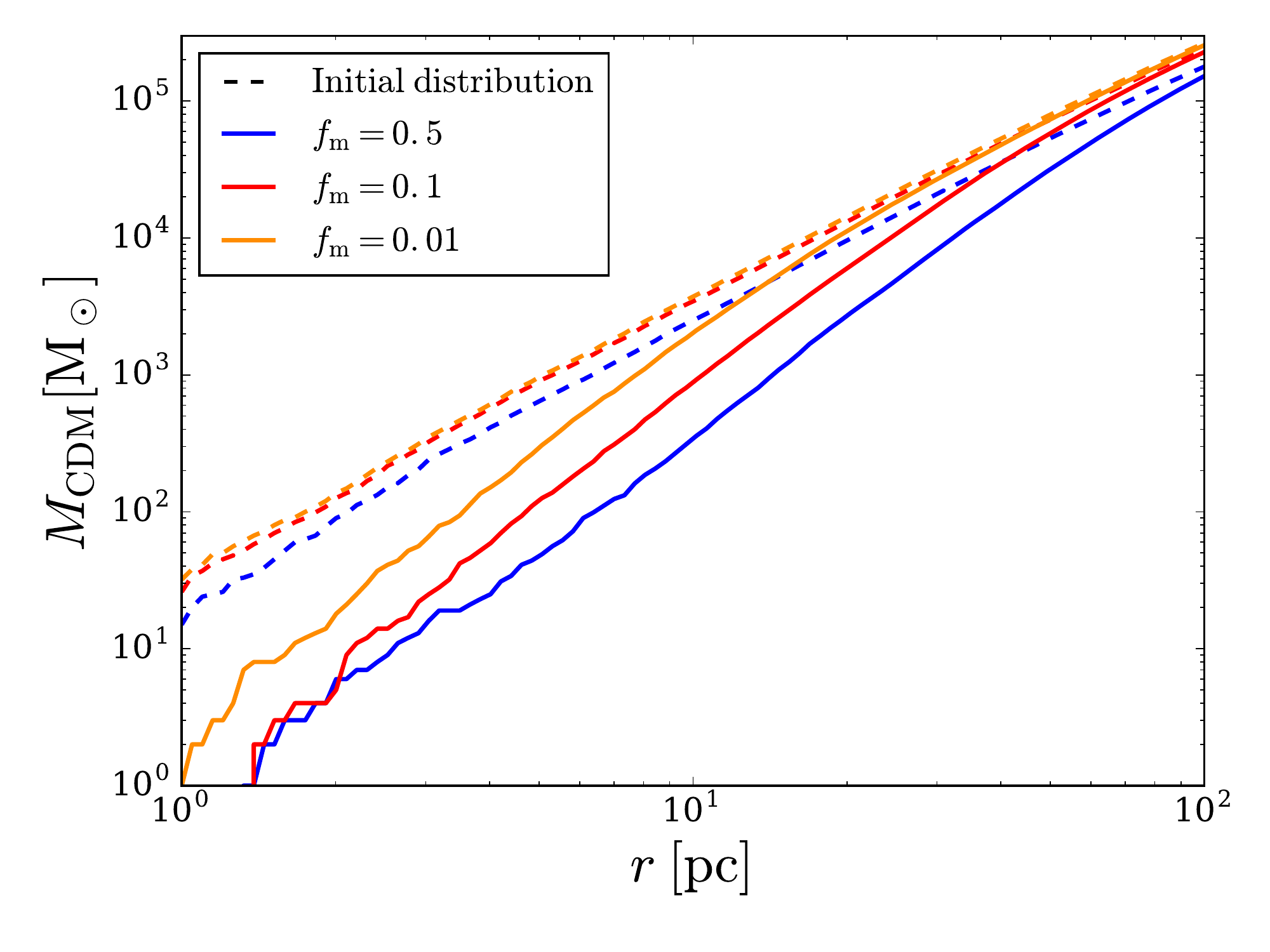}
\end{minipage}
\begin{minipage}[t]{8.4cm} 
\centering
\includegraphics[width=8cm]{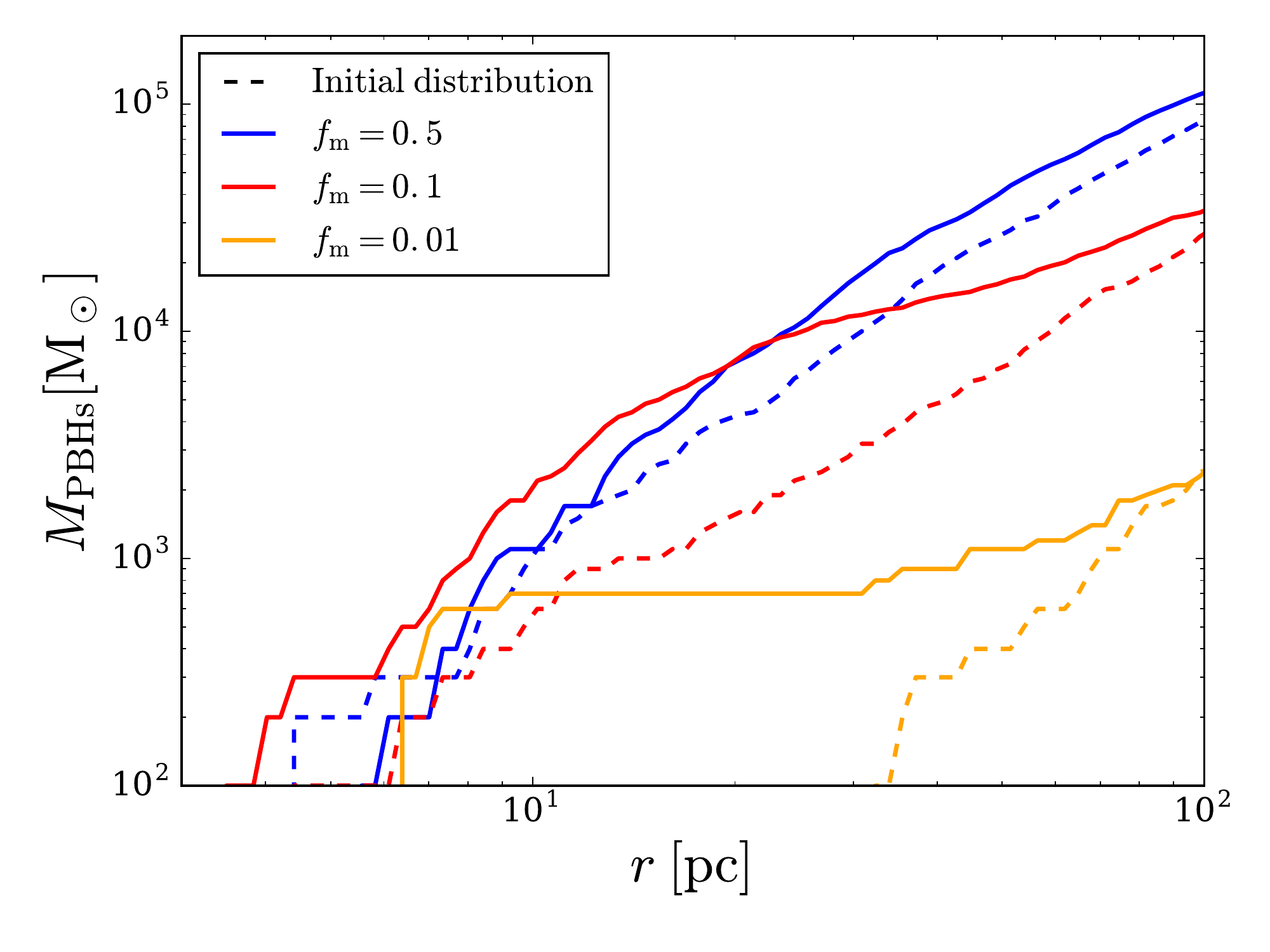}
\end{minipage} 
\caption{{\it Auto-redistribution of CDM particles and PBHs:} Comparison between $T=0$ Gyr (dashed line) and $T=11$ Gyr (solid line) of the interior mass of the CDM component ({\it left panel}) and 100 M$_{\sun}$ PBHs ({\it right panel}) as a function of radius for the three different mass fraction $f_m$ and and $r_{\mathrm{s}}^{\mathrm{PBH}} = r_{\mathrm{s}}^{\mathrm{CDM}}$. The profiles show that CDM particles moved to outer regions and PBHs felt to the halo center. This phenomenon is due to dynamical friction by the CDM environment on PBHs and to a lesser extent by two-body relaxation between PBHs and CDM.}
\label{fga2}
\end{figure*}

\begin{figure}
\centering
\includegraphics[width=0.47\textwidth]{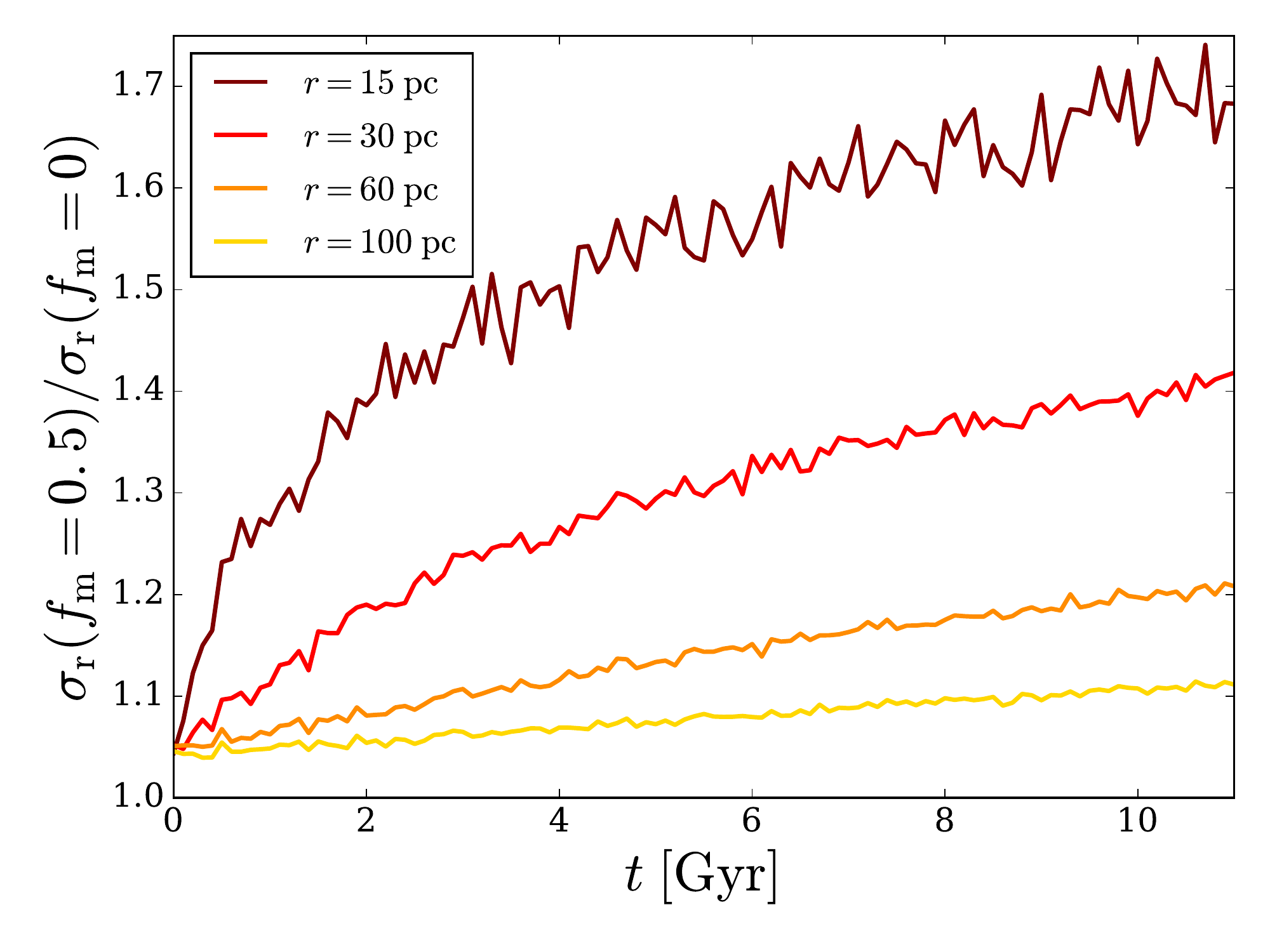}
\caption{{\it Heating of the CDM component:} Ratio between CDM radial velocity dispersion for $f_\mathrm{m}=0.5$ and $f_\mathrm{m}=0$ halos as a function of time at different radii. The PBH+CDM halo with $f_\mathrm{m}=0.5$ contains 100 M$_{\sun}$ PBHs following an initial NFW profile with $r_{\mathrm{s}}^{\mathrm{PBH}} = r_{\mathrm{s}}^{\mathrm{CDM}}$. Over  time, the velocity dispersion ratio increases, especially at the center, due to PBH heating processes, which are two-body relaxation and dynamical friction effect of PBHs. This leads to core formation.}
\label{fga3}
\end{figure}

\begin{figure*}
\centering
\includegraphics[width=\textwidth]{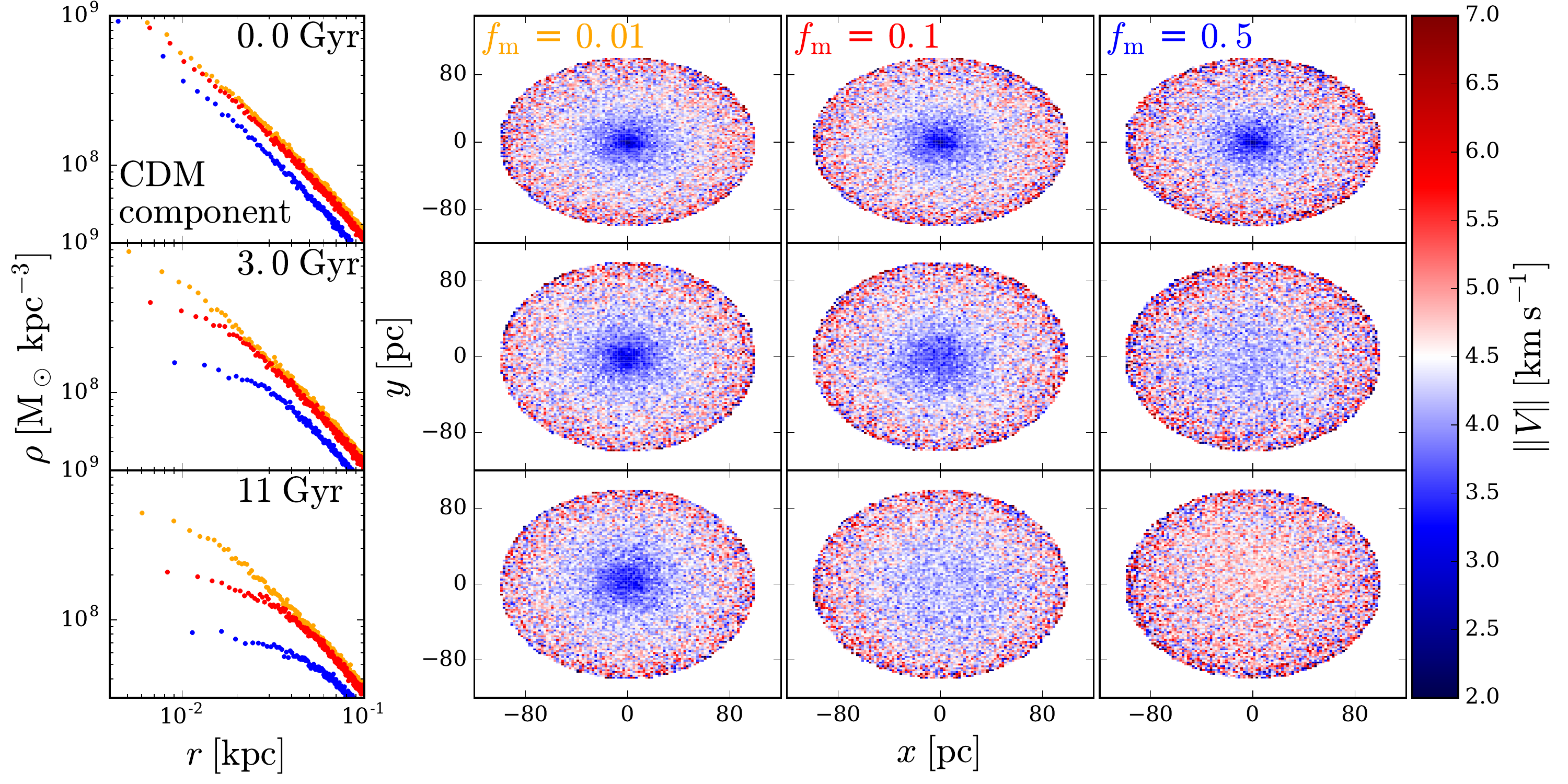}
\caption{{\it Cusp-to-core transition due to heating:} Density profiles of the CDM component ({\it left panels}) for $f_\mathrm{m}=0.01$ (yellow), $0.1$ (red) and $0.5$ (blue) with 100 M$_{\sun}$ PBHs and $r_{\mathrm{s}}^{\mathrm{PBH}} = r_{\mathrm{s}}^{\mathrm{CDM}}$ and their corresponding maps of the CDM mass-weighted velocity distribution projected face-on through a 100 pc ({\it right panels}) at 0, 3 and 11 Gyr. As the CDM velocity increases in the central region, the CDM density profile changes until core formation occurs. All the maps show that core formation goes with dynamical heating of DM particles.}
\label{fga4}
\end{figure*}

First, we assess the impact on the PBH+CDM halo density profiles in our simulations of the two numerical parameters, which are the accuracy control parameter and the softening length $\epsilon$. We tested $2^{-6}$, $2^{-7}$, $2^{-8}$ and $2^{-9}$ for the accuracy control parameter in a PBH+CDM halo with $f_\mathrm{m}=0.1$ and $m_\mathrm{PBH}=100$ M$_{\sun}$. The left panel of the Figure~\ref{fga1} shows that PBH+CDM density profiles hold the initial distribution for all the different values of the accuracy control parameter. Additional accuracy tests confirmed also that the density profile hold the initial distribution for the considered $f_\mathrm{m}$ and $m_\mathrm{PBH}$ values. To test how the softening length impacts on the density profile of the CDM component and the CDM core size, we ran simulations with three different softening lengths $\epsilon$ = $\epsilon_0$/2, $\epsilon_0$ and 2$\epsilon_0$ in order to ensure that our simulations do not suffer from numerical noise. We applied this test on a PBH+CDM halo with $f_\mathrm{m}=0.5$ and $m_\mathrm{PBH}=100$ M$_{\sun}$ expected to have core formation. The middle panel of Figure~\ref{fga1} reveals that softening length (or particle size) does not affect the density profile of CDM component and its core size. Thus, numerical artifacts are not responsible for core formation. We tested also the stability of our halo composed only with CDM particles ($f_\mathrm{m}$=0) over 11 Gyr. We compare our halo profiles at the beginning (T=0 Gyr) and at the end (T=11 Gyr) of the simulation, which are nearly identical for all radii, on the right panel of Figure~\ref{fga1}. We added the density profile of a halo composed only with PBHs ($f_\mathrm{m}$=1), which highlights core formation due to two-body relaxation between PBHs. 

As we show that numerical effects do not initiate core formation, we need to provide evidence for the dynamical mechanism which will induce the cusp-to-core transition. Figure~\ref{fga2} compares the interior mass of the CDM component ({\it left panel}) and 100 M$_{\sun}$ PBHs ({\it right panel}) as a function of radius for three different mass fractions $f_m$ between $T=0$ Gyr and $T=11$ Gyr. Profiles in Figure~\ref{fga2} illustrate that contrary to the CDM particles, the number of PBHs increases in the central region. This results in the PBH infalling towards the central region. By falling in, PBHs will transfer energy to the CDM field via dynamical friction. This is the reason why CDM particles move to the outer regions as we  see in the left panel of Figure~\ref{fga2}. Another important dynamical effect is two-body relaxation between PBHs, which  enhances the CDM particle migration. The CDM velocity dispersion is sensitive to these energy exchanges between CDM particles and PBHs. Figure~\ref{fga3} compared the CDM radial velocity dispersion for $f_\mathrm{m}=0.5$ and $f_\mathrm{m}=0$ halos as a function of time at different distances from the halo center. The PBH+CDM halo with $f_\mathrm{m}=0.5$ contains 100 M$_{\sun}$ PBHs following an initial NFW profile with $r_{\mathrm{s}}^{\mathrm{PBH}} = r_{\mathrm{s}}^{\mathrm{CDM}}$. We highlight that the velocity dispersion ratio increases rapidly, especially in the central region over time, due to PBH heating processes, which are two-body relaxation and dynamical friction effect of PBHs. This leads to core formation. In addition, Figure~\ref{fga4} shows the evolution of the CDM density profiles and the CDM mass-weighted velocity distribution projected face-on through a 100 pc slice during a cusp-to-core transition for the CDM component in a PBH+CDM halo with $f_\mathrm{m}=0.01$, $0.1$ and $0.5$ by assuming 100 M$_{\sun}$ PBHs and $r_{\mathrm{s}}^{\mathrm{PBH}} = r_{\mathrm{s}}^{\mathrm{CDM}}$. As the CDM velocity increases in the central region, the CDM density profile changes until core formation occurs. This figure demonstrates that core formation goes along with dynamical heating of CDM particles.

\subsection{Core size and core formation time}

\begin{figure*}
\centering
\includegraphics[width=\textwidth]{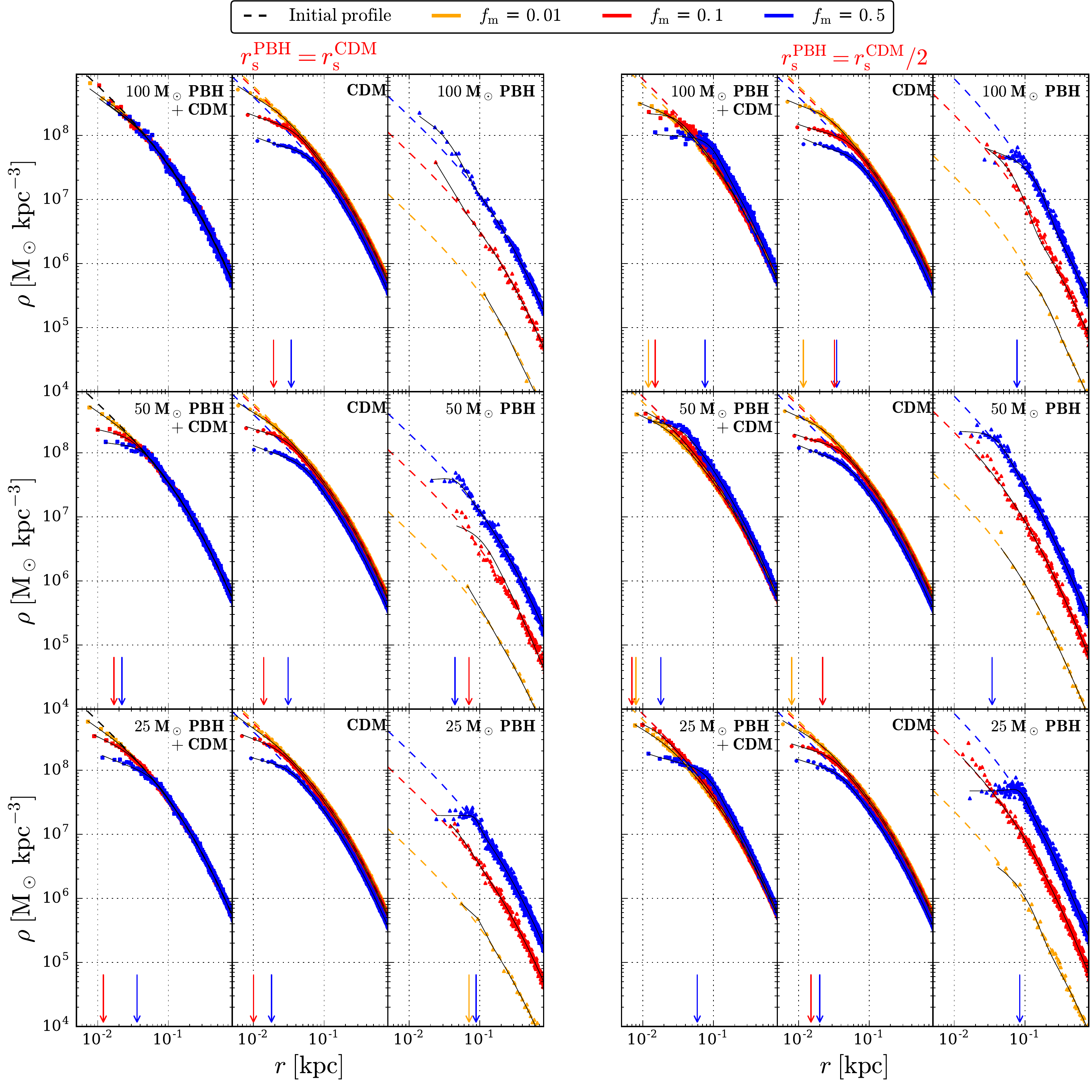}
\caption{{\it Cusp-to-core transitions in PBH+CDM halos:} In each subset, the left panel corresponds to the overall density profiles of the PBH+CDM halo, the middle and right panels to the density profiles of the CDM and PBH components of the PBH+CDM halo separately after T = 11 Gyr. Initially, the PBH and CDM components of the PBH+CDM halo with a total mass of $10^7$ M$_{\sun}$ follow NFW density profiles (dashed lines). We tested two different scale lengths for the profile of the PBH component: $r_{\mathrm{s}}^{\mathrm{PBH}} = r_{\mathrm{s}}^{\mathrm{CDM}}$ (left column) and $r_{\mathrm{s}}^{\mathrm{PBH}} = r_{\mathrm{s}}^{\mathrm{CDM}}/2$ (right column). We explore three different PBH+CDM mass fractions $f_{\mathrm{m}}$ = [0.01, 0.1, 0.5] (see Equation~\ref{hop}). Heating processes between PBH and CDM particles induce a transition from cusp to core in a PBH+CDM halo. In order to determine if there is this transition in our PBH+CDM halo, we did fits (black lines) for the three different density profiles and the size of formed cores are marked by arrows. Our results suggest that it is natural to have multiple cores for a two component halo. Indeed, the core radii of the PBH+CDM halo, the CDM component and the PBH component differ. In the case where the density of PBHs is enhanced in the central region ($r_{\mathrm{s}}^{\mathrm{PBH}} = r_{\mathrm{s}}^{\mathrm{CDM}}/2$), the dynamical heating by PBHs generate larger core sizes (see also Table \ref{T1}).}
\label{fgr1}
\end{figure*}

\begin{table}
\centering
  \label{tab:landscape}
  \begin{tabular}{lccccccc}
    \hline
     $f_\mathrm{m}$ & $r_{\mathrm{s}}^{\mathrm{PBH}}$/$r_{\mathrm{s}}^{\mathrm{CDM}}$ & $m_{\mathrm{PBH}}$ & CCT & $r_{\mathrm{c}}$ & $\chi^2/\nu$\\
     & & [M$_{\sun}$] & & [pc] & \\
    \hline

    0.5 & 1 & 25 & $\checkmark$ & 17.6 & 1.01\\
    0.5 & 1 & 50 & $\checkmark$ & 31.28 & 1.02\\
    0.5 & 1 & 100 & $\checkmark$ & 34.32 & 1.0\\
    
    0.1 & 1 & 25 & $\checkmark$ & 10.01 & 1.01\\
    0.1 & 1 & 50 & $\checkmark$ & 14.48 & 1.0\\
    0.1 & 1 & 100 & $\checkmark$ & 19.33 & 1.03\\
    
    0.01 & 1 & 25 & $\times$ & - &-\\
    0.01 & 1 & 50 & $\times$ & - & -\\
    0.01 & 1 & 100 & $\times$ & -&-\\
    
    0.5 & 1/2 & 25 & $\checkmark$ & 19.7 & 1.05\\
    0.5 & 1/2 & 50 & $\checkmark$ & 22.48 & 0.99\\
    0.5 & 1/2 & 100 & $\checkmark$ & 34.63 & 1.06\\
    
    0.1 & 1/2 & 25 & $\checkmark$ & 15.1 & 1.02\\
    0.1 & 1/2 & 50 & $\checkmark$ & 21.55 & 1.02\\
    0.1 & 1/2 & 100 & $\checkmark$ & 32.23 & 1.01\\

    0.01 & 1/2 & 25 & $\times$ & - &-\\
    0.01 & 1/2 & 50 & $\times$ & -&-\\
    0.01 & 1/2 & 100 & $\checkmark$ & 11.73 & 0.99\\
    \hline
\end{tabular}
\caption{Best fit values for core sizes for the CDM component of the PBH+CDM halo with their corresponding reduced chi-squared $\chi^2/\nu$ for all our simulation scenarios. From left to right, the columns give: the PBH-CDM mass fraction; the scale length for the PBH component; the PBH mass; if or no cusp-to-core transition (CCT) occurs; the core radius of the CDM component; the reduced chi-squared. All these cores are resolved in simulations based on our resolution of about 10 pc when a cusp-to-core transition (CCT) occurs. In the case where the density of PBHs is enhanced in the central region ($r_{\mathrm{s}}^{\mathrm{PBH}} = r_{\mathrm{s}}^{\mathrm{CDM}}/2$), the dynamical heating by PBHs generate larger core sizes (see Figure \ref{fgr1}).}
\label{T1}
\end{table}

\begin{figure}
\centering
\includegraphics[width=0.47\textwidth]{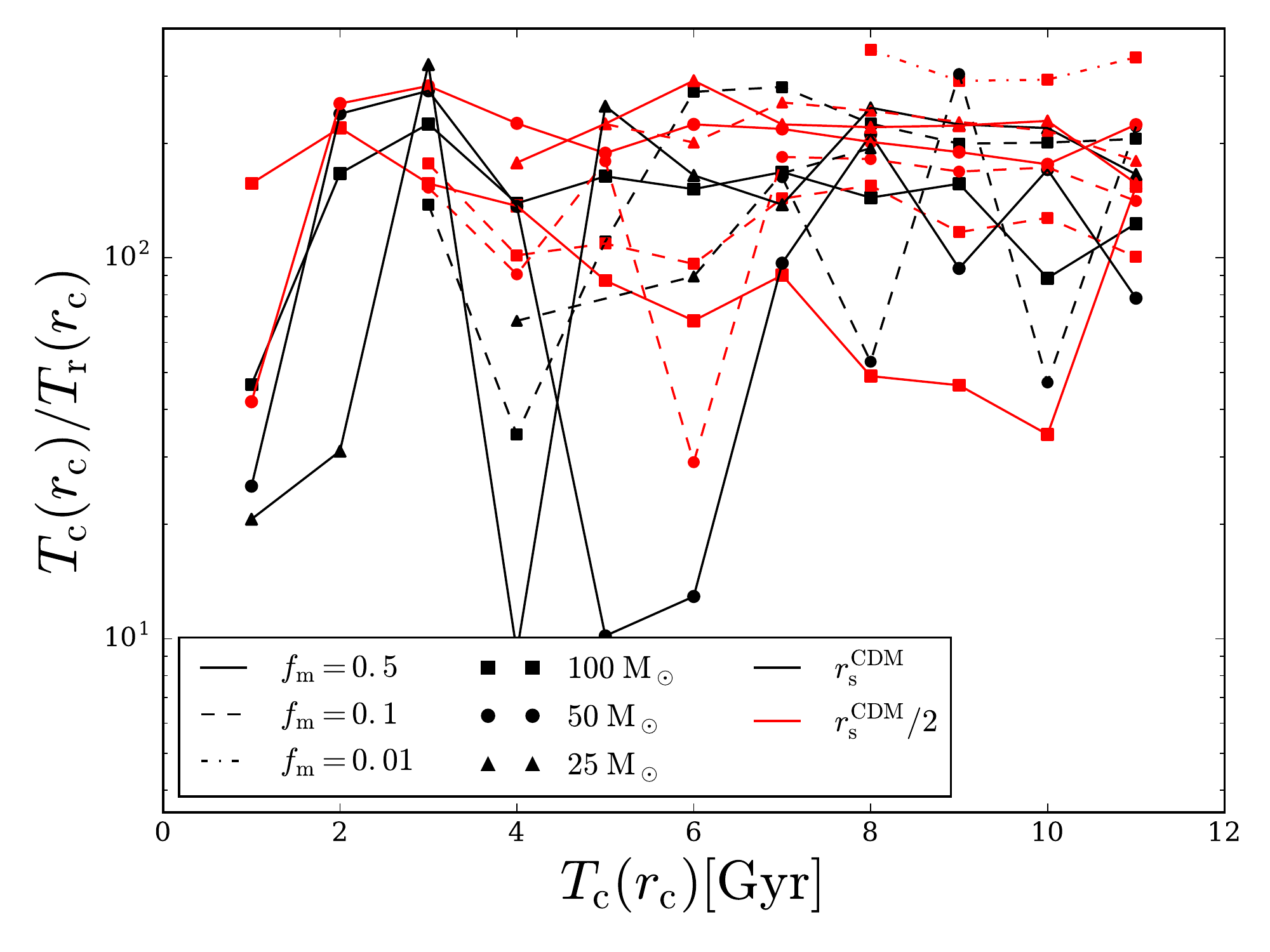}
\caption{{\it Estimating core formation time:} Ratio between the core formation time $T_{\mathrm{c}}(r_{\mathrm{c}})$ and the relaxation time $T_{\mathrm{r}}(r_{\mathrm{c}})$ as a function of the core formation time $T_{\mathrm{c}}(r_{\mathrm{c}})$ for all the simulation scenarios where a cusp-to-core transition occurred (see Table \ref{T1}). This ratio is almost constant over the time in most of simulation scenarios and does not strongly depend on the fraction $f_m$, the PBH mass $m_{\mathrm{PBH}}$ or the PBH scale radius $r_{\mathrm{s}}^{\mathrm{PBH}}$. Indeed, we establish that the time ratio is $\mathcal{O}(100)$ for both scale radii of the PBH component $r_{\mathrm{s}}^{\mathrm{PBH}}$}
\label{fgr12}
\end{figure}

\begin{figure*}
\centering 
\includegraphics[width=\textwidth]{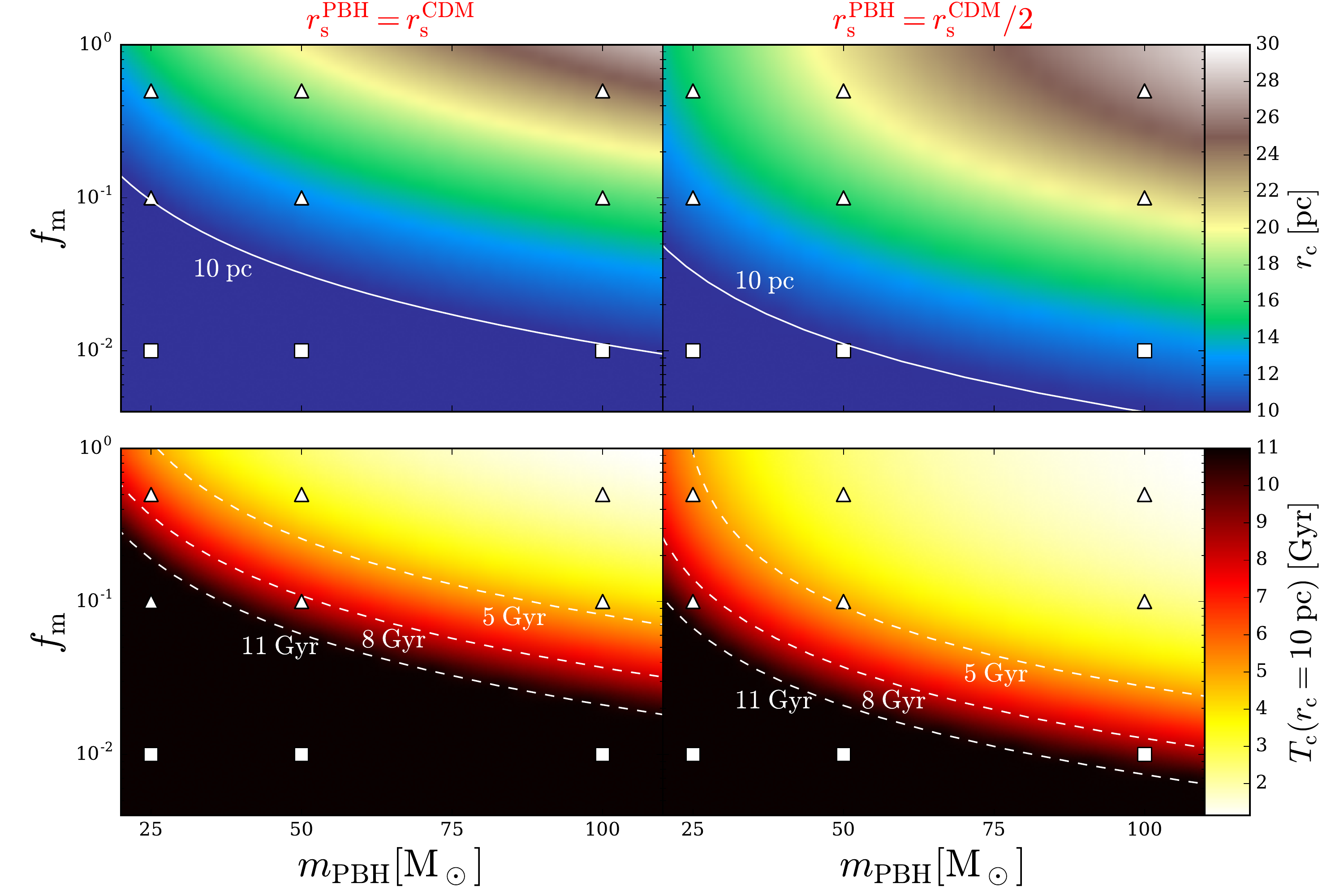}
\caption{{\it Core size and core formation time:} Core size (top panels) and core formation time (bottom panels) maps in ($f_\mathrm{m}$, $m_{\mathrm{PBH}}$) space for both for $r_{\mathrm{s}}^{\mathrm{CDM}}$ models and $r_{\mathrm{s}}^{\mathrm{CDM}}/2$ models. The core radii were calculated for $T_{\mathrm{c}}=$11 Gyr. Triangles (squares) on maps specify that a PBH+CDM halo for a given $f_\mathrm{m}$ and $m_{\mathrm{PBH}}$ has a cored (cuspy) profile based on our simulation results. The white line in top panels, which marks the limit of the cusp-to-core transition, is consistent with the shape of the CDM distribution from the simulations (triangles and squares) under this latter assumption. Core radius maps demonstrates that higher PBH mass and mass fraction in PBH+CDM halo generates larger core sizes. Enhancing the density of PBHs in the central region ($r_{\mathrm{s}}^{\mathrm{PBH}} = r_{\mathrm{s}}^{\mathrm{CDM}}/2$) lowers the threshold of the cusp-to-core transition (white line). At least a mass fraction of $1\%$ is needed to induce cores in PBH+CDM halos depending on the PBH mass and $r_{\mathrm{s}}^{\mathrm{PBH}}$. The core formation time maps (bottom panels) reveal that the cusp-to-core transition takes between 1 and 8 Gyr to occur depending on the fraction $f_m$, the PBH mass $m_{\mathrm{PBH}}$ and the PBH scale radius $r_{\mathrm{s}}^{\mathrm{PBH}}$. The core formation time was calculated at $r_{\mathrm{c}}=$ 10 pc because we assume that a cusp-to-core transition occurred when the core size $r_c$ becomes greater than our spatial resolution of about 10 pc. Enhancing the density of PBHs in the central region accelerate the formation of cores.}
\label{fgr3}
\end{figure*}

Figure \ref{fgr1} shows the overall density profiles of the PBH+CDM halo and the density profiles of the CDM and PBH components of the PBH+CDM halo separately after T = 11 Gyr for different initial mass fractions $f_{\mathrm{m}}$, initial scale radii for the PBH component $r_{\mathrm{s}}^{\mathrm{PBH}}$ and initial PBH masses $m_{\mathrm{PBH}}$. As we expect, two-body relaxation between PBHs and dynamical friction effects from CDM particles modified all the density profiles. In order to determine if there is core formation in our halos, we did a single-component fit for the PBH+CDM profile and separate fits for CDM and PBH profiles. 
We tested many profiles such as Einasto \citep{1965TrAlm...5...87E}, Burkert \citep{1995ApJ...447L..25B} and Zavala (Eq.(4) in \cite{2013MNRAS.431L..20Z}) profiles. However, we found that all of our profiles are well fitted by the following five-parameter formula:
\begin{equation}
    \rho(r) = \rho_{\mathrm{c}} W(r) + (1 - W(r))\rho_{\mathrm{NFW}}(r),
\end{equation}
where $\rho_{\mathrm{c}}$ is the central core density and $W(r)$ is defined as
\begin{equation}
    2W(r)=1-\erf\left(\frac{r-r_{\mathrm{c}}}{2 \Delta}\right),
\end{equation}
where $r_{\mathrm{c}}$ is the core radius and $\Delta$ is a parameter to control the sharpness of the transition from the core to the NFW profile. The error function corresponds to a switching function here. Our fitting formula reproduces the simulated density structures and captures the rapid transition from the cusp to the core. We set Poissonian errors for fitting weights. Our results suggest that it is natural to have multiple cores for a two component halo. Indeed, the core radii of the PBH+CDM halo, the CDM component and the PBH component differ. Based on a reduced chi-squared method, fits of the CDM component of the PBH+CDM halo provide the best-fit values of core radii (see Table \ref{T1}). We consider these core radii to determine whether a transition appears in our simulation. The smallest core size in the simulations is of the order of 10 pc, which corresponds to our spatial resolution. Thus, we assume that a cusp-to-core transition occurred when the core size $r_c$ becomes greater than our spatial resolution. Table \ref{T1} describes the best fit values for core sizes for the CDM component of the PBH+CDM halo for all our simulation scenarios. In the case where the density of PBHs is enhanced in the central region ($r_{\mathrm{s}}^{\mathrm{PBH}} = r_{\mathrm{s}}^{\mathrm{CDM}}/2$), the dynamical heating by PBHs generate larger core sizes (see also Figure \ref{fgr1}).

Now, we have the relation between the core radius $r_c(t)$ of the CDM component of the PBH+CDM halo and the time $t$ when the cusp-to-core transition occurs in the simulations. We invert this function in order to calculate the time ratio $T_{\mathrm{c}}(r_{\mathrm{c}})/T_{\mathrm{r}}(r_{\mathrm{c}})$, where  $T_{\mathrm{c}}(r_{\mathrm{c}})$ and $T_{\mathrm{r}}(r_{\mathrm{c}})$ are the core formation time and the relaxation time, respectively. The relaxation time $T_{\mathrm{r}}$ is given by \citep{2008gady.book.....B}:
\begin{equation}
    T_{\mathrm{r}}(r)\simeq\frac{v^3(r)}{8\pi(n_{\mathrm{CDM}}m_{\mathrm{CDM}}^2 + n_{\mathrm{PBH}}m_{\mathrm{PBH}}^2)G^2\ln{\left(\frac{r_{\mathrm{200}}}{\epsilon}\right)}},
\label{tr}
\end{equation}
where $n_{\mathrm{CDM}}$ and $m_{\mathrm{CDM}}$ ($n_{\mathrm{PBH}}$ and $m_{\mathrm{PBH}}$) are the number-density and mass of CDM (PBH) particles, respectively. $v$, $\epsilon$ and $r_{\mathrm{200}}$  represent the velocity, the softening length and the virial radius, respectively. 

Figure \ref{fgr12} illustrates the time ratio $T_{\mathrm{c}}(r_{\mathrm{c}})/T_{\mathrm{r}}(r_{\mathrm{c}})$ as function of the core formation time $T_{\mathrm{r}}(r_{\mathrm{c}})$ for all the simulation runs where a cusp-to-core transition occurred (see Table \ref{T1}). We notice that the ratio $T_{\mathrm{c}}(r_{\mathrm{c}})/T_{\mathrm{r}}(r_{\mathrm{c}})$ is almost constant over the time in most of simulation runs. Furthermore, it does not strongly depend on the fraction $f_m$, the PBH mass $m_{\mathrm{PBH}}$ or the PBH scale radius $r_{\mathrm{s}}^{\mathrm{PBH}}$. We derive that a cusp-to-core transition occurred for $T_{\mathrm{c}}(r)/T_{\mathrm{r}}(r)\lesssim300$. The discrepancy between the time ratio values is certainly due to our core estimation based only on the CDM component.

Figure \ref{fgr3} represents core size and core formation time maps in ($f_\mathrm{m}$, $m_{\mathrm{PBH}}$) space for both for $r_{\mathrm{s}}^{\mathrm{CDM}}$ models and $r_{\mathrm{s}}^{\mathrm{CDM}}/2$ models. The core radii were calculated for $T_{\mathrm{c}}=$11 Gyr. In order to draw these maps, we predict $r_{\mathrm{c}}$ and $T_{\mathrm{c}}$ by setting the time ratio $T_{\mathrm{c}}(r_{\mathrm{c}})/T_{\mathrm{r}}(r_{\mathrm{c}})$ at 300 and calculating the relaxation time (see Equation \ref{tr}). Indeed, the white line in top panels of Figure \ref{fgr3}, which marks the limit of the cusp-to-core transition, is consistent with the shape of the CDM distribution from simulations (triangles and squares) under this latter assumption. As one can see, core radius maps in Figure \ref{fgr3} demonstrates that higher PBH mass and mass fraction in PBH+CDM halo generates larger core sizes. Enhancing the density of PBHs in the central region ($r_{\mathrm{s}}^{\mathrm{PBH}} = r_{\mathrm{s}}^{\mathrm{CDM}}/2$) lowers the threshold of the cusp-to-core transition (white line). Moreover, our maps show that at least a mass fraction of $1\%$ is needed to induce cores in PBH+CDM halo depending on the PBH mass and $r_{\mathrm{s}}^{\mathrm{PBH}}$. The core formation time was calculated at $r_{\mathrm{c}}=$ 10 pc because we assume that a cusp-to-core transition occurred when the core size $r_c$ becomes greater than our spatial resolution of about 10 pc. These maps (see bottom panels in Figure \ref{fgr3}) reveal that the cusp-to-core transition takes between 1 and 8 Gyr to occur depending on the fraction $f_m$, the PBH mass $m_{\mathrm{PBH}}$ and the PBH scale radius $r_{\mathrm{s}}^{\mathrm{PBH}}$. Enhancing the density of PBHs in the central region accelerate the formation of cores.

We highlighted that the dynamical heating of the CDM component by PBHs can induce cusp-to-core transitions without the presence of baryons and happens automatically in all PBH+CDM halos. Experimentally, annihilation rates from pure CDM halos can prove or prove or disprove this scenario quite clearly\citep{2005Natur.433..389D,2010ApJ...723L.195I}. A baryonic feedback scenario requires starbursts that occur at a particular resonance frequency for a given galaxy potential \citep{2011ApJ...736L...2O}. A single event blowout, results in a temporary core that quickly reverts back to a cusp. This mechanism shown in this work can work even in such failed cases. In addition, low mass galaxies, such as our PBH+CDM halos, can merge after the cusp-to-core transition in order to form more massive galaxies ($10^8$ M$_{\sun}$) with a larger halo cores, which are consistent with observed galaxies. In fact, only a merger of two cored halos yields a cored halo, because a merger of a cuspy halo with a cored halo or a second cuspy halo produces cuspy halo \citep{2004MNRAS.349.1117B}.

\section{Conclusions}

In this paper, we address one of the unresolved problem at small scales, the cusp-core problem, in $10^7$ M$_{\sun}$ halos such as low-mass dwarf galaxies by considering the possibility that a fraction of the DM is made of PBHs. We show that the dynamical heating of the CDM component through PBH infall and two-body relaxation between PBHs induce the formation of cores in PBH+CDM halos. We ran N-body simulations with a high performance and fully GPU-adapted code in order to provide strong evidence for these dynamical heating processes responsible for core formation. We confirm that PBHs as DM candidates can initiate a cusp-to-core transition in these low-mass galaxies. Our results suggest also that it is natural to have multiple cores for a two component halo. Then, we test the PBH+CDM mass fraction $f_\mathrm{m}$ and PBH mass $m_{\mathrm{PBH}}$. We work with PBHs in the 25-100 M$_{\sun}$ mass window, which is consistent with the LIGO detections. Our simulations allow a mass resolution of 1 M$_{\sun}$ for CDM particles. Finally, we derive a criterion based on the relaxation time in order to determine if a cusp-to-core transition occurred: $T_{\mathrm{c}}(r)/T_{\mathrm{r}}(r)\leq300$. Based on our criterion, we set the lower limit on the PBH+CDM mass fraction to be $1\%$ of the total dark matter content to induce cores in PBH+CDM halo depending on the PBH mass and $r_{\mathrm{s}}^{\mathrm{PBH}}$. Here, we have shown that this scenario works even for a small fraction of PBHs. We determine that the cusp-to-core transition takes between 1 and 8 Gyr to appear, depending on the fraction $f_m$, the PBH mass $m_{\mathrm{PBH}}$ and the PBH scale radius $r_{\mathrm{s}}^{\mathrm{PBH}}$. After a transition, the major impact of the PBH+CDM mass fraction and PBH mass is on the core size. Indeed, a larger PBH fraction and PBH mass will induce a larger core radius. As cores occur naturally in PBH+CDM halos without the presence of baryons, there is no cusp-core problem in this alternative theory. As low mass galaxies require less than 8 Gyr to form cores, higher mass galaxies with larger cores as observed can form in the hierarchical scenario. The existence of PBHs in the mass range studied here, 25 - 100 M$_{\sun}$, can possibly be confirmed by the LISA mission.

\section*{Acknowledgments}
We thank the anonymous referee for  helpful comments and suggestions that improved our work. We would like to thank Dante Von Einzbern, Apolline Guillot and Georges Boole for their constructive suggestions to improve the manuscript. YM is supported by ``Joint Usage/Research Center for Interdisciplinary Large-scale Information Infrastructures'' and ``High Performance Computing Infrastructure'' in Japan (Project ID: jh180045-NAH, jh190057-NA).


\bsp	
\label{lastpage}
\end{document}